\documentclass[journal]{IEEEtran}
\usepackage{indentfirst}
\usepackage{changepage}
\usepackage{xcolor}

\usepackage{cite}
\usepackage{enumerate}

\ifCLASSINFOpdf
  \usepackage[pdftex]{graphicx}
  \graphicspath{{../pdf/}{../jpeg/}}
  \DeclareGraphicsExtensions{.pdf,.jpeg,.png}
\else
\fi
\usepackage[cmex10]{amsmath}
\usepackage{tikz}
\usepackage{pgfplots}
\pgfplotsset{width=8cm,compat=1.9}

\usepackage{array}
\newcommand{\PreserveBackslash}[1]{\let\temp=\\#1\let\\=\temp}
\newcolumntype{C}[1]{>{\PreserveBackslash\centering}p{#1}}
 \newcolumntype{R}[1]{>{\PreserveBackslash\raggedleft}p{#1}}
\newcolumntype{L}[1]{>{\PreserveBackslash\raggedright}p{#1}}

\usepackage{textcomp}

\usepackage{booktabs}
\usepackage{threeparttable}
\usepackage{multirow}
\usepackage{caption}
\usepackage{subcaption}
\usepackage{amsmath,amssymb,amsthm}
\usepackage{amsmath, bm}
\usepackage{algorithm}
\usepackage{algorithmic}

\newcommand{\keypoint}[1]{\vspace{0.1cm}\noindent\textbf{#1}}
\newcommand{\cut}[1]{}

\begin{document}
\title{Few-Shot Website Fingerprinting Attack}

\author{Mantun~Chen,
% Michael~Shell,~\IEEEmembership{Member,~IEEE,}
Yongjun~Wang,
Zhiquan~Qin
        % John~Doe,~\IEEEmembership{Fellow,~OSA,}
        % and~Jane~Doe,~\IEEEmembership{Life~Fellow,~IEEE}% <-this % 
        and~Xiatian~Zhu
        % stops a space
\thanks{
M. Chen, Z. Qin and Y. Wang are with College of Computer, National University of Defense Technology, Changsha 410073, China. E-mail: \{chenmantun19,tanzhiquan14,wangyongjun\}@nudt.edu.cn
% M. Shell was with the Department
% of Electrical and Computer Engineering, Georgia Institute of Technology, Atlanta,
% GA, 30332 USA e-mail: (see http://www.michaelshell.org/contact.html).
}
% <-this % stops a space
% \thanks{J. Doe and J. Doe are with Anonymous University.}% <-this % stops a space
\thanks{Xiatian Zhu is with University of Surrey, Stag Hill, University Campus, Guildford GU2 7XH, UK. E-mail: eddy.zhuxt@gmail.com}% <-this % stops a space
% \thanks{Manuscript received April 19, 2005; revised August 26, 2015.}
}


\parindent=19pt

% \markboth{Journal of \LaTeX\ Class Files,~Vol.~14, No.~8, August~2015}%
% \markboth{Journal of \LaTeX\ Class Files,~Vol.~14, No.~8, August~2015}
% {Shell \MakeLowercase{\textit{et al.}}: Bare Demo of IEEEtran.cls for IEEE Journals}



\maketitle

\begin{abstract}
This work introduces a novel data augmentation method
for few-shot website fingerprinting (WF) attack where only a handful of training samples per website are available for deep learning model optimization.
Moving beyond earlier WF methods relying on manually-engineered feature representations, more advanced deep learning alternatives
demonstrate that learning feature representations automatically
from training data is superior.
Nonetheless, this advantage is subject to an {\em unrealistic} assumption that there exist many training samples per website, which otherwise will disappear.
To address this, we introduce a model-agnostic, efficient, and {\em Harmonious Data Augmentation} (HDA) method that can improve deep WF attacking methods significantly.
HDA involves both intra-sample and inter-sample data
transformations that can be used in harmonious manner to expand a tiny training dataset
to an arbitrarily large collection, therefore effectively and explicitly addressing the intrinsic data scarcity problem.
We conducted expensive experiments to validate our HDA for boosting state-of-the-art deep learning WF attack models in both closed-world and open-world attacking scenarios, at absence and presence of strong defense.
{For instance, in the more challenging and realistic evaluation scenario with WTF-PAD based defense, 
our HDA method surpasses the previous state-of-the-art results by more than 4\% in absolute classification accuracy in the 20-shot learning case.}
\end{abstract}


% \fi
\IEEEpeerreviewmaketitle

\section{Introduction}
\begin{figure}[h!]
\centering
\includegraphics[scale=1]{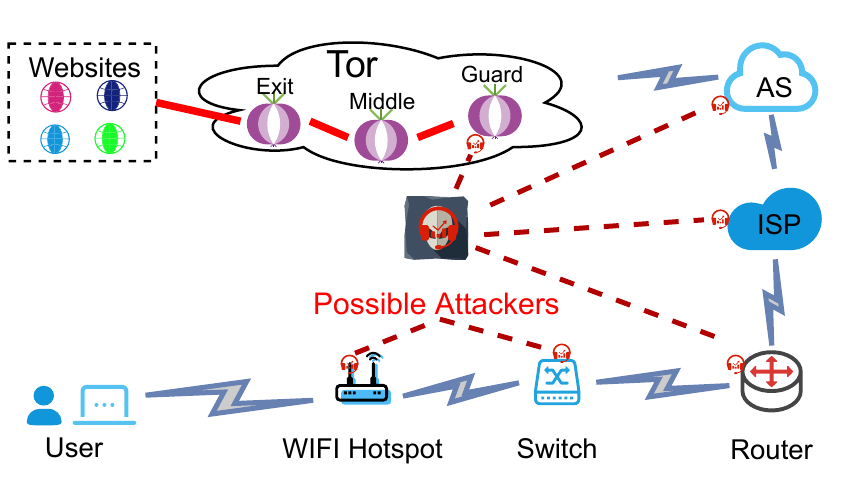}
\caption{Illustration of data flow traffic between a user and 
target websites with a Tor network in-between.
Despite being more secure by anonymity,
website fingerprinting attackers are still able to
reason about which website a victim user is visiting by 
analyzing the data traffic characteristics at multiple locations,
as specified by red dash lines.
% Possible Locations Utilized by Possible Attackers while user Visits Websites by Tor
}
\label{fig:tor_dataflow}
\end{figure}

\IEEEPARstart{F}{or} privacy protection in accessing Internet, an increasing number of users have turned to anonymous networks and Tor \cite{Dingledine2004} is one of the most popular choices \cite{Tor2018}. 
However, this remains not completely secure due to exposure
of data transportation pattern before reaching Tor servers.
For instance, a local attacker would eavesdrop the connection between a user and the guard node of Tor networks, with 
the attacking positions including any devices in the same LAN or wireless network, switch, router, and compromised Tor guard node (see Figure \ref{fig:tor_dataflow}).
By just analyzing the patterns of data package traffic without observing the
content inside, the attacker is likely to reason about which website
a target user is visiting.
This is often known as {\em website fingerprinting} (WF) attack \cite{Hintz2003}.

% In Tor, WF attack allows a local attacker to eavesdrop anywhere of the connection between a user and the guard node of Tor, as shown in Figure 1. The position can be any device in the same LAN or wireless network, switch, router, and compromised Tor guard node.

To implement WF attack, the attacker needs to first create a particular
digital fingerprint for every individual website,
and then learn some intrinsic pattern characteristics of
these fingerprints for accomplishing attack.
Earlier attacking methods rely on
manually designed features based on expert domain knowledge \cite{Hintz2003,Liberatore2006,Bissias2006,Herrmann2009,Panchenko2011,Cai2012,Wang2013,Wang2014,Hayes2016, Panchenko2016}.
They are not only inflexible but also susceptible to environmental changes over time.
This limitation can now be solved by using more advanced deep learning
techniques \cite{LeCun2015}. This is because other than utilizing manually designed features,
deep learning methods can learn automatically feature representations directly from training data, and are hence more scalable provided that 
up-to-date training data are accessible.
A couple of latest state-of-the-art studies, Deep Fingerprinting (DF) \cite{Sirinam2018} and Var-CNN \cite{Bhat2019}, have demonstrated
this potential in comparison to manual feature based methods.
However, these deep learning solutions are not perfect,
as their success is established upon an
unrealistic assumption that a sufficiently large number (e.g. hundreds) of
training samples per website are available.
That is, data hungry.
When only a small training dataset is given as typical in practical use,
their performances are not necessarily superior to traditional methods
\cite{Wang2014,Hayes2016,Panchenko2016}.
% For instance, CUMUL \cite{Panchenko2016}
% \newthing{how to cite? cite for just traditional methods is right, but this is not right for this sentence or this statement}.
%
It is always expensive, tedious, or even infeasible to collect a vast training set in reality due to highly frequent and continuous changes in Internet environments.
Consequently, WF attack is fundamentally a {\em few-shot learning} problem,
which nevertheless is largely unrecognized in the literature.

The nature of few-shot WF attack
is also considered in the recent Triplet Fingerprinting method \cite{Sirinam2019},
under a condition that there is a large set of relevant auxiliary training samples for model pre-training.
It is essentially a transfer learning setting.
This will limit significantly its scalability in practice
as {\em in-the-wild} changes of Internet data traffic conditions
would render such assumptions to be invalid at high probabilities.
On the contrary, we introduce a realistic, generic few-shot WF attack setting
where only a handful of training samples are available for every target website,
without making any domain-specific assumptions.
Clearly, Triplet Fingerprinting is not applicable in our setting
due to the need of auxiliary training data.

We summarize the \textbf{contributions} of this paper as follows:
\begin{description}
\item{\bf (I)}
We introduce a novel, practical {\em few-shot website fingerprinting attack} problem,
in which only a few training samples are available without rich 
auxiliary data.
This respects the intrinsic nature
of highly dynamic Internet traffic conditions 
and high cost of collecting large training data in practice.
Highlighting the importance of {\em few-shot learning} without any
auxiliary data assumption for the first time,
we hope more future efforts would be dedicated for solving 
practically important WF attack challenge. 
\newline
\item{\bf (II)}
To solve the proposed few-shot learning challenges,
we embrace the enormous potentials and advantages of deep learning
for WF attack by introducing a new {\em Harmonious Data Augmentation} (HDA) method
to explicitly solve the training data scarcity problem in deep learning.
Specifically, we augment the original training data
by rotating and masking-out randomly individual samples
and mixing (linearly combining) sample pairs in arbitrary proportions.
With such intra-sample and inter-sample data transformations,
our HDA method can efficiently expand a tiny training dataset
at any scales.
\newline
\item{\bf (III)}
We benchmark the performance of
few-shot WF attack
and demonstrate the efficacy of our data augmentation method
using existing state-of-the-art deep learning models.
In particular, we consider 5-20 shots per website/class
in closed-world and open-world settings,
with and without defense.
The results show that our method can improve the performances of previous state-of-the-art deep learning solutions
\cite{Bhat2019, He2016} significantly.
\end{description}
\section{Related Work}

\subsection{Objectives, Scenarios and Assumptions}
The objective of WF attack is to identify which website a victim user is interacting with among a set of monitored target websites.
Conceptually, it is a multi-class classification problem
with each website regarded as a unique class.
There are several scenarios with different assumptions. The most common scenario is \textit{closed-world} attack,
where the user is assumed to only visit a set of known target websites under monitoring. 
This assumption however is not realistic, therefore
discarded in the \textit{open-world} scenario.
In this scenario, the victim user is considered to likely visit any websites including those monitored ones, as typically experienced in real-world applications. 

A third scenario considers \textit{defense} where the user takes some actions to defend against potential attack.
This would lead to higher attack difficulty. 
Representative defense techniques include Buflo \cite{Dyer2012}, Tamaraw \cite{Cai2014}, Walkie-Talkie \cite{Wang2017} and WTF-PAD \cite{Juarez2016}. 
Among them, WTF-PAD is used as the mainstream method for Tor 
networks due to low bandwidth overhead and zero delay. 
We considered WTF-PAD based defense in our evaluations.

In the literature, several common assumptions are made.
We briefly discussed three main assumptions. 
In \textit{user behavior}, it is assumed that all Tor users browsed websites sequentially, only opening a single tab at a time.
In \textit{background traffic}, it is assumed that the attacker is able to collect all the clean traces generated by the victim's visits against dynamic background traffic.
This is increasingly possible as shown in \cite{Wang2016}
the multiplexed TLS traffic can be split into individual encrypted connections to each website. 
In \textit{network condition}, 
the attacker is assumed to have the same conditions as the victim including traffic condition and settings.
To compare with the benchmark results, we follow these general assumptions for fair evaluations.

Instead, we focus on addressing the following assumption.
Often, the attacker assumes that the training data fall into 
a similar distribution as the deployment data.
This is a particularly strong and artificial assumption
as the network condition is actually changing and evolving frequently.
Such a property enforces the attacker to update the training data in order to have a robust attacking model over time.
This implies that the attacker is less possible to collect a large set of training data at each time due to high acquiring costs.
However, existing WF attack methods often ignore this factor
by assuming availability of large training data.
In contrast, we study the largely ignored few-shot learning setting in WF attack. 
Specifically, we approach this problem by explicitly solving
the small training data issue via synthesizing new labelled training data.

\subsection{Website Fingerprinting Attack Methods}

The first pioneer attack against Tor networks was evaluated by Herrmann \cite{Herrmann2009} in 2009. It achieved an accuracy of 2.96\% using around 20 training samples per website in the closed-world scenario. 
Later on, Wang and Ian \cite{Wang2013} proposed to represent the traffic data using more fundamental Tor cells (i.e., direction data) as a unit rather than TCP/IP packets.
This representation is rather meaningful and informative as
it encodes essential characteristics of Tor data.
By training a SVM classifier with distance-based kernel, a ground-breaking performance with 90.9\% accuracy was achieved on 100 sites each with 40 training samples.
Recently, Panchenko et al. \cite{Panchenko2016} proposed an idea of sampling the features from a cumulative trace representation
and achieved 91.38\% accuracy with 90 training instances per website. 
Hayes and Danezis \cite{Hayes2016} exploited random decision forests to achieve similar results.
% in the same setting.
A common design of these above methods is 
a two-stage strategy including feature design and classifier
learning. 
This is not only constrained by the limitations of hand features
but also lacks interaction between the two stages,
making the model performance inferior.

Motivated by the remarkable success of deep learning techniques in computer vision and natural language processing \cite{krizhevsky2012imagenet,collobert2011natural},
several deep learning WF attack methods have been introduced
which can well solve the aforementioned weakness.
This is because deep learning methods by design carry out
feature learning and classification optimization from the raw training data end-to-end.
For example, using VGG network \cite{simonyan2014vgg} as the backbone, Sirinam et al. \cite{Sirinam2018} proposed a Deep Fingerprinting attack (DF) model that attains 98.3\% accuracy on 95 websites. 
However, this method needs a large training set (e.g. 1000 training samples per website), otherwise it will suffer from 
significant performance drop.
When using 20 training samples per website,
DF can only hit around 19.4\% accuracy.
To overcome this limitation, Bhat et al. \cite{Bhat2019} developed the Var-CNN model based on ResNet \cite{He2016} and dilated causal convolution \cite{DenOord2016,yu2016multi-scale}.
When small training sets (e.g. 100 samples per website)
are available, it achieves superior performance over DF
but at dependence on less realistic time features
and less scalable hand-crafted statistical information.
% However, it relies on extra

A solution to few-shot learning is a recently proposed triplet fingerprinting (TF) method \cite{Sirinam2019}.
The key idea of TF is to pre-train a metric model 
that can measure pairwise distances on new classes.
%as feature extractor,
% TF further deploys a triplet network for few-shot learning.
% Building upon a pre-trained DF model as feature extractor,
% TF further deploys a triplet network for few-shot learning.
When the pre-training dataset is similar to the target data
in distribution, TF can hit an accuracy of 94.5\% on 100 websites using only 20 training samples per website.
This is a strong transfer learning scenario.
However, considering that the dynamics of network conditions
is highly unknown and uncontrollable, 
such a transfer learning assumption is hardly valid in practice.
In light of this observation, in this work we propose a more realistic few-shot learning setting without assuming
any auxiliary data with similar data characteristics for model pre-training.
Hence 
it is more scalable and generic for real-world deployments.
Under the proposed more challenging few-shot setting,
TF is unable to work properly due to insufficient network initialization.

\subsection{Data Augmentation}
Data augmentation is an important element in deep learning
due to its data-hungry nature \cite{LeCun2015}.
For example, random insertion, random swap, and random deletion
for text classification in natural language processing \cite{Wei2019EDA},
or geometric transformations (e.g., flipping, rotation, translation, cropping, scaling), color space transformations (e.g., color casting, varying brightness, and noise injection),
inter-image mixup \cite{Zhang2018}
for image analysis \cite{Morenobarea2018NoiseInjection,Taylor2017, shorten2019a}.
These previous attempts have shown the significance of
different augmenting methods for model performance 
on the respective tasks.
Inspired by these findings, we investigate extensively
the effectiveness of training data augmentation 
by adapting existing operations for deep learning WF attack in few-shot learning settings. To the best of our knowledge,
this is the first attempt of its kind.
Crucially, we demonstrate that existing {state-of-the-art} deep WF attack method   \cite{Bhat2019} significantly benefits from 
using the proposed data augmentation operations
in varying evaluation scenarios.
This result would be encouraging and influential for future investigation 
of deep learning WF attack methods in particular.

\section{Method}
\subsection{Problem Definition}
In website fingerprinting (WF) attack, 
the {\em objective} is to detect which website a target user is visiting.
The common observations are data traffic traces $\bm{x}$
produced 
by one visit to a website $y$.
Taking each website as a specific class, this is essentially a multi-class classification problem.
For model training, a 
labelled training set $D = \{(\bm{x}_i, y_i)\}_{i=1}^N$
is often provided,
where $y_i \in \{1,2,\cdots,K\}$
specifying one of $K$ target websites.
Two different settings are often considered in model testing:
(1) {\em Closed-world} attack where 
any test sample is assumed to belong one of target websites/classes,
and 
(2) {\em Open-world} attack where
the above assumption is eliminated,
i.e., a test trace may be produced by a {\em non-target} (unmornitored) website.
The latter is a more realistic setting,
yet presenting a more challenging task
as identifying if a test sample falls into target classes or not
is non-trivial.

\keypoint{Feature representation. }
For common Tor networks, the raw representation of a specific traffic trace 
consists of a sequence of temporally successive Tor cells 
% that form the traffic 
travelling between a target user and a website visited. 
It is derived from TCP/IP data.
Specifically,
after those TCP/IP packets retransmitted are discarded,
TLS records are first reconstructed, their lengths
are then rounded down to the nearest multiple of 512
to form the final sequence data $\bm{x}$.
In value, each $\bm{x}$ is a sequence of 1 (outgoing cell) and -1 (incoming cell),
with a variable length.
This raw representation is hence known as
direction sample.
Besides, temporal information about inter-packet time
is another modality of data used, but limited by
high reliance on network conditions, 
i.e., not stable and much more noise.
Consequently, we mainly consider the direction data samples
in this study, which are more scalable and generic.

\subsection{Deep Learning for Website fingerprinting Attack}
\begin{figure}[h!]
\centering
\includegraphics[scale=1]{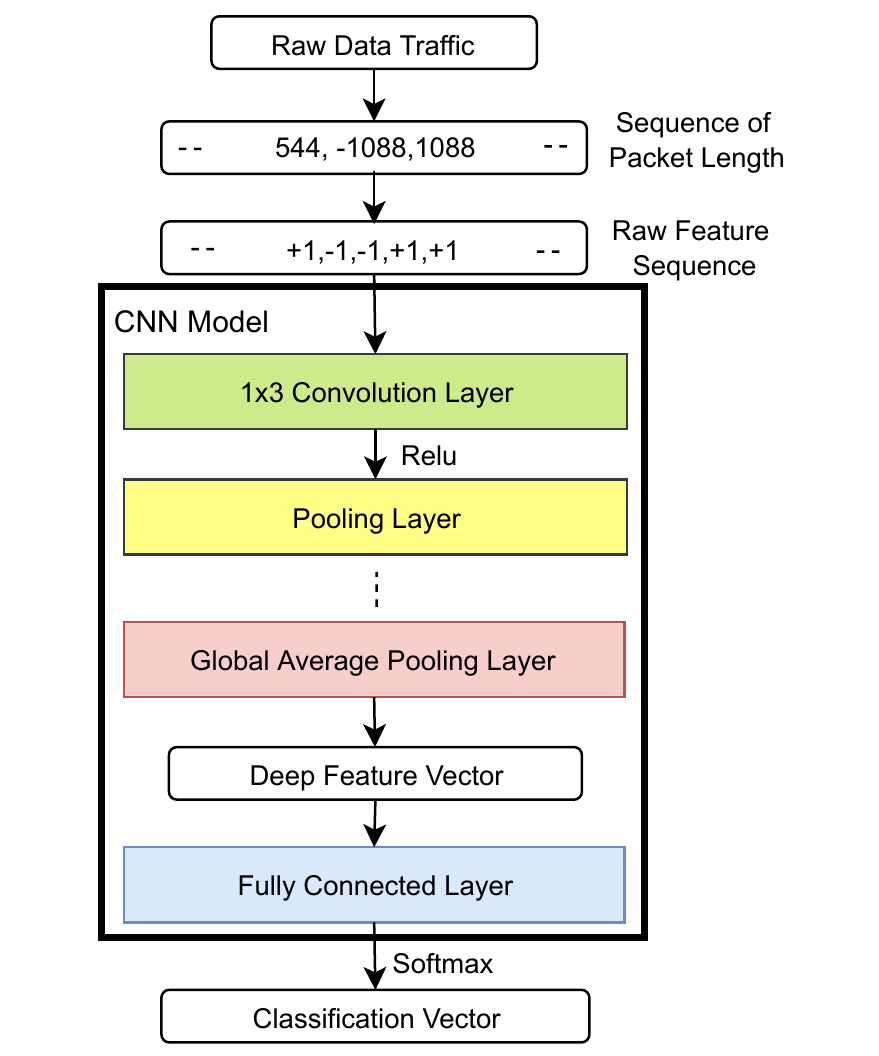}
\caption{A deep learning pipeline for website fingerprinting attack that conducts
feature representation and website classification end-to-end in a joint learning manner.
% \note{TODO for TUN: Just show a deep learning pipeline:
% input is raw feature sequences, then deep feature vectors, lastly classification result.
% Please connect to text.
% }
}
\label{fig:deep_net}
\end{figure}

Most of existing WF attack methods 
rely on hand-crafted feature representations
\cite{Hintz2003,Liberatore2006,Bissias2006,Herrmann2009,Panchenko2011,Cai2012,Wang2013,Wang2014,Hayes2016, Panchenko2016}.
This strategy is not only unscalable but also unsatisfactory in performance
due to limited and incomplete domain knowledge.
Deep learning methods provide a viable solution
via learning directly more effective and expressive
representation from training data,
as shown in a few recent studies \cite{Sirinam2018, Bhat2019}.
In this work we advance this new direction further.
% \newthing{my checking stop here.}

We explore 1D convolutional neural networks (CNN) \cite{kiranyaz2015convolutional}
for WF attack as the raw data are temporal sequences.
Building on the success of deep learning 
in computer vision,
we adopt the same high-level network designs
of standard 2D CNN models \cite{Lecun1998}
whilst translating them into 1D counterparts.
This is similar to \cite{Sirinam2018,Bhat2019}.

A CNN model consists of multiple convolutional layers with non-linear activation functions such as ReLU \cite{agarap2018ReLU} and fully-connected (FC) layers, 
characterized by end-to-end feature extraction and classification. 
With convolutional operations, the filters of each layer 
transform input sequences using learnable parameters and output new feature sequences. 
This feature transformation is conducted layer by layer in a hierarchical fashion.
The receptive field (kernel) with size 3 is often used in each individual layer to capture local feature patterns. 
By stacking more layers and pooling operation, 
the model can perceive the information
of larger regions and achieve translational invariance.
Another effective method for enlarging receptive field
is dilated causal convolutions \cite{DenOord2016,yu2016multi-scale},
which has been exploited in \cite{Bhat2019}.

The feature representations $\bm{f}$ of WF samples
are the output of global average pooling layer
on top of the last convolution layer.
To obtain the classification probability vector $\hat{\bm{y}} = \{\hat{y}_1, \hat{y}_2, \cdots, \hat{y}_K \} \in \mathcal{R}^K$ over $K$ target classes,
$\bm{f}$ is fed into a FC layer and normalized by
a softmax function.

For model training, we compute a cross-entropy objective loss function with the classification vector
against the ground-truth class label
over all $N$ training samples as: 
\begin{equation}
   \mathcal{L} = \sum_{i=1}^N \sum_{j=1}^K \delta(j=y_i) \log \hat{y}_{i,j} 
   \label{eq:CE}
\end{equation}
where $y_i$ refers to the ground-truth class label of a training sample $\bm{x}_i$, and $\delta()$ is a Dirac function.
The objective is to maximize the probability
of the ground-truth class in prediction.
This loss function is differentiable,
with its gradients backpropogated to update all the learnable model parameters.

Once the deep model is trained, 
we forward a given test sample,
obtain a classification probability vector,
and take the most likely class as prediction
in both closed-world and open-world settings.
For open-world setting, 
all unmornitored websites are considered
to belong to a background class.

\keypoint{Discussion. }
While deep learning techniques have advanced
significantly in the last several years,
it is still assumed that a large set of labelled training samples is available. 
This is not always true, for example,
for the WF attack problems.
In real-world applications,
an attacker is usually faced with highly dynamic 
network environments.
It means that the distribution of raw features 
is evolving continuously.
As such, the training data need to update frequently,
which disables collection of large training data with labels in practice
due to prohibitively high labelling costs.
Consequently, only a small training set is 
accessible in reality,
making deep learning methods ineffective.

\subsection{Harmonious Website Fingerprinting Data Augmentation}
To address the above small training data challenge,
we propose an intuitive, novel {\em harmonious data augmentation} (HDA) method.
We introduce both {\em intra-sample} and {\em inter-sample} 
augmentation operations that can be applied in a joint and harmonious manner for more effective 
data expansion.

\keypoint{Intra-sample augmentation. }
The key idea of intra-sample augmentation
is that given an individual training sample, we introduce a certain degree of {\em random} data perturbation and/or variation whilst keeping the 
same class labels.
Doing so allows us to generate an infinite number
of labelled training samples due to the nature of randomness.
We consider two perturbation operations:
random rotation and random masking.

\begin{figure*}[h!]
\centering
\includegraphics[width=0.89\linewidth]{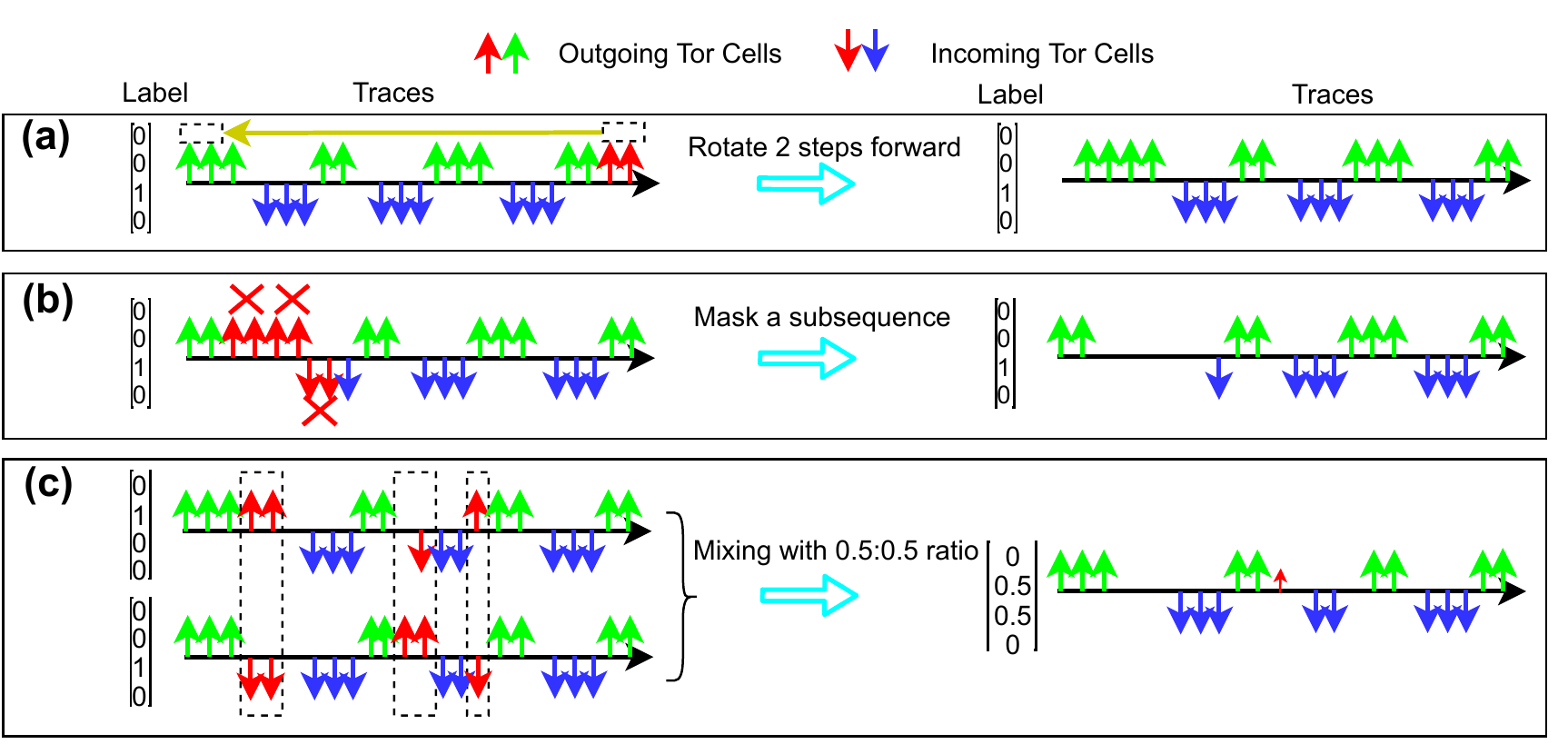}
\caption{Illustration of our data augmentation operations for deep learning WF attack, including (a) random rotation, (b) random masking, and (c) random mixing.}
\label{fig:data_aug}
\end{figure*}

{\em Random rotation} based data augmentation
means rotating an original training sample forward or backward by random steps to generate virtual
samples (Fig. \ref{fig:data_aug}(a)):
% between 0$^{\circ}$ and 359$^{\circ}$. 
\begin{equation}
    Rotate(\bm{x}, \; n_\text{step}, \; \text{dir})
\end{equation}
where $n_\text{step}$ and $\text{dir} \in \{\text{forward, backward}\}$ specify the steps and the direction to rotate
on an input sample $\bm{x}$. 
The hypothesis behind is that
class-sensitive information encoded in a sample
is distributed across different sub-sequences
and data traffic order is less important 
than signal patterns.
After a sampled is rotated, the original class information is largely preserved, i.e., semantically invariant.
Hence, the same class can be annotated for the rotated variants.
However, this hypothesis is more likely
to stand under some certain (unknown) degrees.
We therefore introduce an upper bound parameter
$R_\text{max}$ so that
the rotation range is limited at most $R_\text{max}$
steps in both directions,
$n_\text{step} \leq R_\text{max}$.

In contrast, {\em random masking} introduces
localized corruption to an original training sample
by setting a random subsequence to zero (Fig. \ref{fig:data_aug}(b)).
This data augmentation is written as:
\begin{equation}
    Mask(\bm{x}, \; n_\text{len}, \; \text{loc})
\end{equation}
where $n_\text{len}$ and $\text{loc}$
denote the length and location of the subsequence
that is masked out from an original sample $\bm{x}$.
Rather than in form of subsequence, 
another strategy is to randomly select
individual positions to mask.
We consider this
% \newthing{this? the latter ? which is better?}
may introduce more significant
corruption to the underlying semantic information.

Conceptually, random masking simulates varying traffic measurement errors
in data transportation. 
Meanwhile, with the same above hypothesis,
such masking would not dramatically change the
semantic class information, provided that
the masking is subject to some limit,
e.g., the length of subsequences masked out $M_\text{len}$.
It hence offers a complementary data perturbation choice
w.r.t. random rotation.

\keypoint{Inter-sample augmentation. }
Apart from data augmentation on individual samples,
we further introduce data perturbation across two different samples
to enrich the limited training set.

We propose {\em random mixing} that generates virtual samples and class labels
by linear interpolation between two original samples
$\bm{x}_{i}$ and $\bm{x_{j}}$ as:
\begin{align}
% \tag{1}
&\bm{\tilde{x}} = \lambda \bm{x}_{i} + (1- \lambda)\bm{x}_{j}\\
&\bm{\tilde{y}} = \lambda \bm{y}_{i} + (1- \lambda)\bm{y}_{j}
\end{align}
% \end{flalign*}
where % $(\bm{x}_{i},\bm{x_{j}})$ are traces and 
$(\bm{y}_{i},\bm{y_{j}})$ are the one-hot class labels of 
$\bm{x}_{i}$ and $\bm{x_{j}}$. 
The mixing parameter $\lambda \in [0, 1]$ 
follows a Beta distribution: 
$\lambda \sim \beta(\alpha,\alpha)$ with $\alpha > 0$ the parameter
that controls the strength of interpolation.
This is in a similar spirit of mixup in image understanding domain \cite{Zhang2018}.
Unlike intra-sample augmentation above, random mixing
changes the semantic class information, % of original samples,
since original samples may be drawn from different classes.
It simplifies the data distribution by
imposing a linear relationship between classes
for complexity minimization.
While seemingly counter-intuitive,
we will show that such a method brings positive 
contributions on top of random masking and random
rotation.

\keypoint{Combination and compatibility. }
Different augmentation operations can be applied
on the same samples without conflict to each other in a harmony.
There is also no particular constraint
on the order of applying all the three data augmentation operations in a combination.
Given a fixed set of parameters as discussed above,
different augmentation orders will result in
different virtual samples.
This makes little conceptual difference
as the space of sample is just infinite.

\keypoint{Augmentation optimization. }
In our harmonious data augmentation (HDA), three hyper-parameters $\{R_\text{max}, M_\text{len}, \alpha\}$ are introduced.
To generate meaningful virtual samples,
obtaining their optimal values is necessary
otherwise adversarial effects may even be imposed.

Instead of manual tuning, we adopt an automatic Bayesian estimator,
called Tree of Parzen Estimators (TPE) \cite{RN702}.
The conventional TPE can take only one parameters alone at a time.
So we need to optimize each of the three hyper-parameters independently. 
This differs from our data augmentation process
where the three augmentation operations are typically applied together,
making the independently tuned parameters of TPE
% \newthing{tuning each parameter independently?} 
sub-optimal.
This is because, jointly applying three augmentations together makes them inter-dependent.

For solving this problem,
we propose a sequential optimization process
that takes into account the inter-dependence property of different augmentation operations gradually (see Alg. \ref{alg}).
Specifically, we start with a random, fixed order
of applying our random rotation, masking, and mixing operations.
Then, we optimize from the first one with TPE, 
move to next one with all the previous ones optimized and fixed, stop by finishing the last one.
Each time, we still optimize a single hyper-parameter whilst keeping all the previous optimized ones fixed. In this way, we expand the inter-dependence among different operations sequentially.

\renewcommand{\algorithmicrequire}{\textbf{Input:}}
\renewcommand{\algorithmicensure}{\textbf{Output:}}
\begin{algorithm}[!h]
	\caption{Data augmentation optimization}%算法标题
	\begin{algorithmic}[1]%一行一个标行号
		\REQUIRE 
		A training $\{\bm{X_{t}}, \bm{Y_{t}}\}$,
		and validation $\{\bm{X_{v}}, \bm{Y_{v}}\}$ set.
		\\ [0.2cm]
        \ENSURE Data augmentation with optimal parameters $B_\text{aug}$.
        \\ [0.2cm]
        \STATE Setting $B_\text{aug}$ = $\phi$ (empty set);
        \STATE Sequencing data augmentation operations randomly; \\ [0.1cm]
        \WHILE{Enumerating augmentation operations}
		\STATE Get the search space $S_\text{aug}$ of current augmentation $A$;
		\STATE Using TPE on $S_\text{aug}$ to obtain
		the optimal parameter $b_\text{aug}$,
		with the model trained by $B_\text{aug}$ and $A$;
		\STATE $B_\text{aug} = B_\text{aug} \cup b_\text{aug}$
		\ENDWHILE \\ [0.1cm]
		\RETURN $B_\text{aug}$
	\end{algorithmic}
\label{alg}
\end{algorithm}

\subsection{Theoretical Foundation and Formulation}

The objective of learning a WF attack model
is equivalent to derive a function $h \in H$ that fits the
latent translation relationship between raw feature vectors
$\bm{x} \in X$ and corresponding website class labels $\bm{y} \in Y$.
That is, fitting a joint distribution $P(X, Y)$.
To this end, in deep learning we often leverage 
a loss function $L$ defined to penalize the differences between predictions $h(\bm{x})$ and targets $\bm{y}$. 
We minimize the average loss over the joint distribution: 
\begin{equation}
    R(h) = \int L(h(\bm{x}), \bm{y}) \; dP(\bm{x},\bm{y})
\end{equation}
which is known expected risk minimization\cite{Vapnik1998}.

However, the joint distribution is often unknown,
particularly for WF attack with small training data.
Given a limited training data set $D = \{ (\bm{x}_i, \bm{y}_{i}) \}_{i=1}^N$, 
the joint distribution can only be approximated by an empirical distribution as:
\begin{equation}
% \tag{3}
P_{\delta}(\bm{x}, \bm{y}) = \frac{1}{N} \sum^{N}_{i=1}\delta(\bm{x} = \bm{x}_{i}, \bm{y} = \bm{y}_{i})
\end{equation}
where $\delta(\bm{x} = \bm{x}_{i}, \bm{y} = \bm{y}_{i})$ is a Dirac mass centered at a sample $(\bm{x}_{i}, \bm{y}_{i})$. Accordingly, the expected risk can now be approximated by an empirical risk:
\begin{align}
% \tag{4}
R_{\delta}(h) & = \int L(h(\bm{x}), \bm{y})d P_{\delta}(\bm{x},\bm{y}) \\ \nonumber
& =\frac{1}{N} \sum^{N}_{i=1}L(h(\bm{x}_{i}), \bm{y}_{i})
\end{align}
The above approximation is in the empirical risk minimization (ERM) principle \cite{Vapnik1998}. 
The cross-entropy loss (Eq. \eqref{eq:CE})
is a representative example,
which essentially minimizes $R_{\delta}(h)$ 
for the classification task.

While ERM is a common strategy,
it suffers from high risk of poor generalization
due to the tendency of memorization,
particularly when a large model is used \cite{Szegedy2013Intriguing}.
To mitigate this issue,
we adopt the notion of vicinal distribution \cite{Chapelle2000} which can better approximate the true joint distribution.
In particular, the vicinal distribution $P_{\upsilon}$ in the data space is defined as:
\begin{equation}
P_{\upsilon}(\bm{\tilde{x}}, \bm{\tilde{y}}) = \frac{1}{n} \sum^{n}_{i=1}\upsilon(\bm{\tilde{x}_{i}}, \bm{\tilde{y}_{i}}\mid \bm{x}_{i}, \bm{y}_{i})
\end{equation}
Intuitively, $P_{\upsilon}$ measures the probability of finding a virtual labelled sample $(\bm{\tilde{x}}, \bm{\tilde{y}})$ in the vicinity around an original training sample $(\bm{x}_{i}, \bm{y}_{i})$. 

Given such vicinal distributions, we first construct a virtual dataset $D_{\upsilon} := {(\tilde{\bm{x}}_{i}, \tilde{\bm{y}}_{i})}^{m}_{i=1}$ by sampling $P_{\upsilon}$ randomly, and then minimize an empirical vicinal risk to learn $h$ as:
\begin{equation}
R_{\upsilon}(h) = \frac{1}{m} \sum^{m}_{i=1}L(h(\bm{\tilde{x}}_{i}), \bm{\tilde{y}}_{i})
\end{equation}
Clearly, at the core of this strategy is performing
data augmentation around original training samples.
Rather than computing a loss value for every single training 
sample, it derives a local distribution centered at each individual sample and generates more virtual training samples to 
reduce the negative memorization effect of deep learning
This is the key rationale of our data augmentation method.

\keypoint{Augmentation formulation. }
We formulate the proposed harmonious data augmentation operations
in the vicinal distribution manner.
For intra-sample augmentation (including random rotation and masking), the vicinal distribution is defined as
\begin{equation}
% \tag{7}
\upsilon(\bm{\tilde{x}}, \bm{\tilde{y}}\mid \bm{x}, \bm{y}) = T(\bm{x}) \delta(\bm{\tilde{y}} = \bm{y})
\end{equation}
where $T()$ is a transformation operator. 
% Now the problem is how to get the matrix $A$. 

For {\em random rotation}, 
given any length-$n$ sample $\bm{x} = \{x_{0}, ...,x_{i}, ..., x_{n-1}\}$, we first define a circle matrix $B$ for forward rotation
as:
\begin{equation}
\label{eq:rot_mat}
% \tag{8}
    B (\bm{x})
    =     % 第一部分
    \left[
    \begin{smallmatrix}
      x_{0} & x_{1} & \cdots & x_{n-1} \\
      x_{n-1} & x_{0} & \cdots & x_{n-2} \\
      \vdots & \vdots &  & \vdots \\
      x_{n-1} & x_{n-2} & \cdots & x_{0} \\
    \end{smallmatrix}
    \right]       % 第二部分
\end{equation}
Then we sample the step size $n_\text{step}$
{\em uniformly} from a range of $\{1, \cdots, R_\text{max}\}$. 
By one-hot representation of $n_\text{step}$,
we can obtain a rotation transformation as:
\begin{equation}
  T_\text{rot}(\bm{x}) = \text{one-hot}(n_\text{step}) B(\bm{x})
\end{equation}
For the backward case, we perform the same process as above
but with a backward rotation matrix instead.

For {\em random masking}, we similarly sample the start position $s$ {\em uniformly} in the range of $\{1, \cdots, n - n_\text{len}\}$
where $n_\text{len}$ is the length of masked subsequence.
The masking transformation can be represented by a matrix as:
\begin{equation}
% \tag{12}
M_\text{mask} = diag \Big(\bm{1} - \sum^{s+n_\text{len}}_{i=s}{Row_{i}(I)} \Big)
\end{equation}
where $I$ is identity matrix, 
$\bm{1}$ is all-one vector,
$Row_i()$ selects
the $i$-th row of a matrix,
$diag()$ transforms a vector to a diagonal matrix.
Masking operation is finally conducted 
by matrix multiplication as:
\begin{equation}
T_\text{mask}(\bm{x}) = \bm{x} M_\text{mask}
\end{equation}
For inter-sample augmentation, {\em random mixing} in our case, the vicinal distribution is defined as:
\begin{align}
% \tag{14}
& \upsilon(\bm{\tilde{x}}, \bm{\tilde{y}}\mid \bm{x}_{i}, \bm{y}_{i}, \bm{x_{j}}, \bm{y_{j}}) =
\\ \nonumber
& \delta \big(\bm{\tilde{x}} =
\lambda \cdot \bm{x}_{i} + (1 - \lambda) \cdot \bm{x_{j}},
\;\;
\bm{\tilde{y}} = \lambda \cdot \bm{y}_{i} + (1 - \lambda) \cdot \bm{y_{j}} \big)
\end{align}
where $\lambda$ is a random variable
drawn from a Beta distribution
$\beta(\alpha, \alpha)$
% {\note{$$\bm{y}$ is one-hot format?}}
and $\bm{y}$ is one-hot class label vector.
This local vicinity is assumed to 
respect a linear structure w.r.t.
class labels.
\section{Experiments}

\subsection{Experimental Setup}

\keypoint{Datasets. }
We evaluated our data augmentation method HDA
on four standard WF attack datasets as below.
(1) \textbf{CW$_{100}$} \cite{Rimmer2018}:
% For the close-world, t
This dataset provides a total of 100 monitored target websites each with 2,500 raw feature traces.
(2) \textbf{DF$_{95, Nodef}$} \cite{Sirinam2018}:
This dataset gives 95 monitored websites with each 
contributing 1,000 feature traces.
(3) 
\textbf{ROWUM$_{400,000}$} \cite{Rimmer2018}:
This dataset includes CW$_{100}$ and a large set of 
samples each was generated by a visit to 
a page of top 400,000 Alexa websites.
(4) \textbf{DF$_{95,wtf-pad}$} \cite{Sirinam2018}:
Unlike all the above datasets, this is a more challenging dataset
due to presence of WTF-PAD based defense against WF attack. 
It has the same size as DF$_{95,Nodef}$,
i.e., 95,000 raw feature samples from 95 websites.
We considered both closed-world and open-world WF attack
scenarios using the above datasets.

\keypoint{Network architectures. }
We used four different network architectures for 
testing the generic benefits of the proposed HDA method.
(1) Var-CNN \cite{Bhat2019} is the current state-of-the-art
deep learning WF method.
(2) Var-CNN$^*$ is an improved variant with two more
fully-connected layers added to the classifier.
(3/4) ResNet-18 and ResNet-34 \cite{He2016} are 
two strong and popular networks widely deployed in many different fields such as computer vision.

% To evaluate the efficacy of our method (HDA),
% we used the same neural network 
% as deep competitors for fair comparison.

\keypoint{Implementation details. }
We conducted our experiments in Keras \cite{Keras2015}.
In all our experiments, we used the standard training, validation, and test splits for all competitors for fair comparisons.
We optimized HDA's hyper-parameters using Var-CNN \cite{Bhat2019} as deep learning model 
on $CW_{100}$ in 
closed-world setting and applied the same parameter setting for 
all the other deep learning methods, datasets and settings.
This allows testing the generality and scalability
of our HDA method.
For augmentation optimization,
we set the search space as:
$1\sim20$ with step 5 for forward/backward $R_\text{max}$ (random rotation)
$1\sim200$ with step 20 for $M_\text{len}$ (random masking),
$[0, 1]$ with step 0.1 for $\alpha$ (random mixing).
The optimal parameter values we obtained are
$R_\text{max}=20$, 
$M_\text{len}=180$,
and 
$\alpha=0.1$.
For saving storage, we performed online data
augmentation within each mini-batch without 
any data pre-processing.
In each experiment, we trained every deep learning model
for 150 epochs and used the checkpoint with best
performance on the validation
set for model test.
We only used the direction feature data, without 
time sequences and hand-crafted features.
%as in original Var-CNN.
%
We run each experiment by 10 times and reported the mean results and standard deviation as the final performance.

\renewcommand{\arraystretch}{1.5} %控制行高
\begin{table} %[!tp]
  \centering
  \setlength{\tabcolsep}{2pt}
  \caption{Results of {\em closed-world} WF attack on CW$_\text{100}$. Metrics: Accuracy.
  }
  \label{tab:cls_world_CW}
  \resizebox{1\linewidth}{!}{
    \begin{tabular}{c|cccc}
    %{p{1cm}<{\centering} p{1.4cm}<{\centering} p{1.4cm}<{\centering} p{1.4cm}<{\centering} p{1.4cm}<{\centering} }
    \toprule
    Method & 5-shot & 10-shot & 15-shot&20-shot\cr
    \midrule
    CUMUL \cite{Panchenko2016} & $72.2\pm1.7$ & $79.7\pm1.4$ & $83.3\pm2.0$ & $85.9\pm0.6$\cr
    k-FP \cite{Hayes2016} & \boldmath{$79.3\pm1.0$} & $83.9\pm1.0$ & $85.9\pm0.6$ & $87.5\pm0.8$ \cr
    \hline
    % DF \cite{Sirinam2018} & $73.8\pm1.5$ & $84.9\pm1.0$ & $89.4\pm0.7$& $90.1\pm0.5$\cr

    ResNet-18 \cite{He2016} & $13.9\pm0.6$ & $23.0\pm1.1$ & $35.5\pm2.5$& $51.3\pm1.7$\cr
    ResNet-34 \cite{He2016} & $14.5\pm0.7$ & $24.3\pm1.5$ & $40.3\pm3.1$& $51.3\pm86.4$\cr
    Var-CNN \cite{Bhat2019} &$17.9\pm1.5$  & $41.4\pm4.0$ & $65.6\pm1.9$ & $78.7\pm1.5$\cr
    Var-CNN$^{*}$ \cite{Bhat2019} &$34.8\pm2.6$  & $57.4\pm3.9$ & $71.9\pm2.4$ & $80.8\pm2.4$\cr
    \hline
    % DF+{\bf HDA} & $77.9\pm0.5$ & $86.5\pm0.5$ & $90.6\pm0.4$& $92.5\pm0.4$\cr
    % Var-CNN+{\bf HDA} & $77.6\pm2.5$ & \boldmath{$90.4\pm0.9$} & \boldmath{$93.5\pm0.3$} & \boldmath{$94.8\pm0.6$}
    ResNet-18+{\bf HDA} & $34.3\pm4.3$ & $61.5\pm9.4$ & $77.6\pm5.8$& $84.6\pm4.3$\cr
    ResNet-34+{\bf HDA} & $34.8\pm6.2$ & $62.3\pm8.1$ & $78.8\pm7.1$& $86.4\pm2.8$\cr
    Var-CNN+{\bf HDA} & $59.7\pm1.5$ & $74.7\pm2.6$ & $86.4\pm1.3$ & $90.7\pm0.8$\cr
    Var-CNN$^{*}$+{\bf HDA} &$71.3\pm4.7$  & \boldmath{$90.2\pm0.6$} & \boldmath{$93.3\pm0.2$} & \boldmath{$94.1\pm0.5$}\cr
    \bottomrule
    \end{tabular}
    }

\end{table}

\begin{table} %[!tp]
  \centering
  \setlength{\tabcolsep}{2pt}
  \caption{
  Results of {\em closed-world} WF attack on DF$_{95,Nodef}$. Metrics: Accuracy.
  }
  \label{tab:cls_world_DF}
  \resizebox{1\linewidth}{!}{
    \begin{tabular} {c|cccc}
    %{p{1cm}<{\centering} p{1.4cm}<{\centering} p{1.4cm}<{\centering} p{1.4cm}<{\centering} p{1.4cm}<{\centering} }
    \toprule
    Method&5-shot&10-shot&15-shot&20-shot\cr
    \midrule
    % DF \cite{Sirinam2018} & $71.5\pm0.6$ & $82.9\pm0.7$ & $86.1\pm0.7$& $89.0\pm0.9$\cr
    ResNet-18 \cite{He2016} & $14.3\pm1.0$ & $26.3\pm1.6$ & $41.1\pm1.9$& $54.3\pm1.1$\cr
    ResNet-34 \cite{He2016} & $16.3\pm1.3$ & $29.5\pm4.7$ & $41.4\pm3.2$& $54.2\pm2.6$\cr
    Var-CNN \cite{Bhat2019} &$22.6\pm3.4$  & $50.0\pm2.1$ & $75.6\pm1.5$ & $75.2\pm1.7$\cr
    Var-CNN$^{*}$ &$19.4\pm1.6$  & $29.6\pm2.7$ & $42.5\pm5.0$ & $53.6\pm5.2$\cr
    % Var-CNN \cite{Bhat2019} &$0.8\pm0.3$  & $19.4\pm14.6$ & $50.5\pm7.0$ & $73.6\pm1.7$\cr
    \hline
    % DF+{\bf HDA} & \boldmath{ $75.9\pm0.4$} & $84.7\pm0.5$ & $87.9\pm0.7$& $90.4\pm0.4$\cr
    % Var-CNN+{\bf HDA} & $70.8\pm3.4$ & \boldmath{$85.7\pm0.6$} & \boldmath{$88.9\pm0.7$} & \boldmath{$91.7\pm0.6$}\cr
    ResNet-18+{\bf HDA} & $53.9\pm7.0$ & $77.3\pm3.6$ & $82.8\pm2.9$& $88.3\pm1.1$\cr
    ResNet-34+{\bf HDA} & $44.0\pm6.8$ & $74.3\pm7.4$ & $84.2\pm3.1$& $88.7\pm1.6$\cr
    Var-CNN+{\bf HDA}&$71.0\pm2.3$  & $84.3\pm1.4$ & $88.6\pm0.9$ & \boldmath{$91.4\pm0.5$}\cr
    Var-CNN$^{*}$+{\bf HDA} & \boldmath{$62.3\pm4.8$} & \boldmath{$84.3\pm0.9$} & \boldmath{$87.7\pm0.7$} & $91.0\pm0.5$\cr
    \bottomrule
    \end{tabular}
    }
    % \end{threeparttable}
    % \label{table I}
\end{table}

\subsection{Closed-World WF Attack}
\label{subsec:closed_world_exp}

\keypoint{Setting. } We conducted the closed-world attack
on CW$_{100}$ and DF$_{95,Nodef}$.
We separated each dataset into 
training, validation (10 samples per class), 
and test (70 samples per class) splits.
We considered few-shot settings with 
$n \in \{5,10,15,20\}$ training samples per class.
The validation set was used to select
the best performing model for test.
We used the classification accuracy as performance metric.
Besides deep network models, we also compared our method with two conventional hand-crafted feature based methods: CUMUL \cite{Panchenko2016} and k-FP \cite{Hayes2016}.

% We compared with the state-of-the-art deep learning WF method  and its variation(Var-CNN \cite{Bhat2019} and Var-CNN$^{*}$ which have two  additional  dense  layers  in  its  classifier), two other deep learning WF methods(ResNet-18 and ResNet-34 \cite{He2016}), 
% and two conventional methods (CUMUL \cite{Panchenko2016} and k-FP \cite{Hayes2016}).
% We only used the direction feature data, without 
% using time sequences and hand-crafted features as in original Var-CNN.
% To evaluate the efficacy of our method (HDA),
% we used the same neural network 
% as deep competitors for fair comparison.
%
 
\renewcommand{\arraystretch}{1.5} %控制行高
\begin{table*}[!tp]
  \centering
  \setlength{\tabcolsep}{15pt}
  %\fontsize{6.5}{8}\selectfont
%   \begin{threeparttable}
  \caption{Results of {\em open-world} WF attack on CW$_\text{100}$ (target classes) + ROWWUM$_\text{400,000}$ (non-target classes).
  Pre: Precision,
  Rec: Recall.
  We reported two settings: one is tuned for 
  best precision (top), and one for recall (bottom).
  }
  \label{tab:open_world}
%   \resizebox{1\linewidth}{!}
  {
    \begin{tabular}{c cc cc cc cc}
    \toprule
    \multirow{4}{*}{Method}&\multicolumn{8}{c}{Tuned for Precision}\cr
    &\multicolumn{2}{c}{5-shot}&\multicolumn{2}{c}{10-shot}&\multicolumn{2}{c}{15-shot}&\multicolumn{2}{c}{20-shot}\cr
    \cmidrule(lr){2-3} \cmidrule(lr){4-5} \cmidrule(lr){6-7} \cmidrule(lr){8-9}\cr
    &Pre&Rec&Pre&Recall&Pre&Rec&Pre&Rec\cr
    \midrule
    ResNet-18 \cite{He2016} & 28.7&0.8  &30.0&7.6  &49.8&8.7 &65.9&18.3\cr
    ResNet-34 \cite{He2016}& 32.5&4.7  &42.1&6.6  &50.4&21.8 &61.3&39.6\cr
    Var-CNN \cite{Bhat2019} & 39.7&2.7  &58.8&9.2  &74.2&35.8 &78.0&54.4\cr
    Var-CNN$^{+}$&44.7&3.9  &58.9&17.9 &64.0&49.9 &66.1&65.3\cr
    ResNet-18+{\bf HDA} & 80.7&0.2  &89.6&7.2  &$\mathbf{91.7}$&30.7 &$\mathbf{93.4}$&46.6\cr
    ResNet-34+{\bf HDA} & 72.7&0.8  &$\mathbf{91.7}$&8.7  &91.5&43.8 &92.6&55.1\cr
    Var-CNN+{\bf HDA} & 77.4&6.9  &91.2&47.2  &91.4&64.3 &92.9&66.6\cr
    Var-CNN$^{*}$+{\bf HDA}&$\mathbf{92.3}$&5.4  &87.8&59.7 &88.7&73.0 &90.5&76.1\cr
    \toprule
    \multirow{4}{*}{}&\multicolumn{8}{c}{Tuned for Recall}\cr
    \midrule
    ResNet-18 \cite{He2016} &10.0& 19.8  &17.0&34.0  &24.8&49.5 &32.3&64.3\cr
    ResNet-34 \cite{He2016} &12& 25.6  &19.2&37.8  &30.2&54.5 &34.9&68.6\cr
    Var-CNN \cite{Bhat2019} & 13.9&27.1  &26.7&52.8  &37.4&73.9 &42.2&82.6\cr
    Var-CNN$^{+}$&16.3&31.9  &27.6&54.8 &37.0&73.4 &41.1&81.4\cr
    ResNet-18+{\bf HDA} & 20.0&36.4  &36.3&71.5  &44.5&87.4 &47.2&91.0\cr
    ResNet-34+{\bf HDA} & 21.1&37.4  &36.2&68.9  &45.7&89.2 &47.2&91.4\cr
    Var-CNN+{\bf HDA} & 28.0&$\mathbf{55.8}$  &46.9&88.7  &49.3&92.2 &49.1&92.5\cr
    Var-CNN$^{*}$+{\bf HDA}&25.6&48.1  &45.7&$\mathbf{89.3}$ &47.9&$\mathbf{92.8}$ &47.8&$\mathbf{94.1}$\cr
    \bottomrule
    \end{tabular}
    }
\end{table*}

\keypoint{Results. }
The results of different methods are compared
in Table \ref{tab:cls_world_CW} and Table \ref{tab:cls_world_DF}.
We have the following observations:
(1) Hand-crafted feature based methods (CUMUL and k-FP)
remain competitive, especially very few samples per class are available. The best 5-shot result on CW$_\text{100}$
is achieved by k-FP, with a moderate edge of 8\% over Var-CNN$^{*}$+HDA.
(2) However, deep learning methods (Var-CNN$^{*}$) 
becomes clearly stronger
when a few more training samples are accessible,
suggesting a great deal of potentials.
Among previous methods, in 20-shot case Var-CNN$^{*}$ achieves the best result on CW$_\text{100}$ and DF$_{95, Nodef}$.
(3) With our HDA method for training data augmentation,
every deep learning method improves significantly.
For example, the 5-shot accuracy of  Var-CNN is increased from 17.9\% to 59.7\% on CW$_\text{100}$,
and 22.6\% to 71.0\% on DF$_{95, Nodef}$. % \newthing{incredibly?}. 
Similarly, the accuracy of ResNet-18 is improved from 13.9\% to 34.3\% on CW$_\text{100}$,
and from 14.3\% to 53.9\% on DF$_{95, Nodef}$. 
Var-CNN and its variant benefit incredibly more, implying a higher
demand for larger training data to avoid model overfit.
Similar effects are shown for 10/15/20-shot cases.
(4) Our HDA can consistently improve different methods
on varying datasets, suggesting good generality.
(5) The performance deviation of Var-CNN and its variant assisted by our method HDA 
is the least among all the competitors,
implying strong stability.

\subsection{Open-World WF Attack}
\keypoint{Setting. }
We conducted the open-world attack experiments
on the combination of ROWWUM$_\text{400,000}$ and CW$_\text{100}$.
We treat the websites of CW$_\text{100}$
as target (monitored) classes,
and those of ROWWUM$_\text{400,000}$ 
as non-target (unmonitored) classes.
In this test, we selected randomly 8,020 out of 400,000 unmonitored websites,
and separated them into three disjoint sets sized at
20/1,000/7,000 for 
training, 
validation, 
and test, respectively. 
In this scenario, the precision and recall rates were used to evaluate model performance due to the need for detecting non-target classes \cite{Juarez2014}.
We considered the same four deep learning methods (ResNet-18, Resnet-34, Var-CNN \cite{Bhat2019} and its variant Var-CNN$^{*}$) for comparisons.

\keypoint{Results. }
The results of different methods are reported 
in Table \ref{tab:open_world}.
We considered two settings,
one is tuned for best precision,
and one for best recall.
Overall, we obtained similar trends
as above that our HDA is highly effective
for improving both deep learning methods.
It is noted that unlike the closed-world scenario,
VarCNN$^{*}$+HDA achieves very top result at most cases under both tuning settings even if it may not be the best one.
Similarly, VarCNN$^{*}$+HDA remains to be more stable and less sensitive to training sample size.
Importantly, our HDA method further enhances this strengths
by efficient data augmentation,
leading to a more robust WF attack solutions.

\subsection{WF Attack against Defense}
\keypoint{Setting. }
In contrast to the two above experiments, we further tested a more challenging WF attack scenario 
with defense involved.
Defense changes the data traffic patterns to be more similar to each other, therefore making the attack more difficult.
We considered the most popular defense, WTF-PAD,
widely deployed in Tor networks.
We used the DF$_\text{95,wtf-pad}$ dataset in this experiment.
We used 100 random samples per website,
and separated them into three sets for training (20 samples), validation (10 samples), and test (70 samples), respectively.
We reported the classification accuracy as 
performance metric
in closed-world scenario.
We compared with previous four deep learning 
(ResNet-18, Resnet-34, Var-CNN \cite{Bhat2019} and its variant Var-CNN$^{*}$) and 
hand-crafted feature based methods
(k-NN \cite{Wang2014},
SDAE \cite{Abe2016},
k-FP \cite{Hayes2016},
CUMUL \cite{Panchenko2016},
AWF \cite{Rimmer2018}).

\renewcommand{\arraystretch}{1.5} %控制行高
\begin{table}[!tp]
  \centering
  \setlength{\tabcolsep}{2pt}
  \caption{
  Results of {\em closed-world} WF attack with {\em WTF-PAD based defense} on DF$_{95, wtf-pad}$. Metrics: Accuracy.
  }
  \resizebox{1\linewidth}{!}{
  \label{tab:result_defense}
    \begin{tabular} {c|cccc}
    \toprule
    Method&5-shot&10-shot&15-shot&20-shot\cr
    \midrule
    
    k-NN \cite{Wang2014} &--&--&--&16.0\cr
    SDAE \cite{Abe2016} &--&--&--&36.9\cr
    k-FP \cite{Hayes2016} &--&--&--&57.0\cr
    CUMUL \cite{Panchenko2016} &--&--&--&60.3\cr
    AWF \cite{Rimmer2018} &--&--&--&60.8\cr
    ResNet-18 \cite{He2016} & $7.3\pm0.3$ & $9.8\pm0.6$ & $11.4\pm0.4$& $14.2\pm0.6$\cr
    ResNet-34 \cite{He2016} & $7.4\pm0.5$ & $9.4\pm0.7$ & $13.3\pm1.3$& $12.3\pm1.2$\cr
    Var-CNN \cite{Bhat2019} &$6.6\pm0.3$  & $9.2\pm0.7$ & $12.5\pm0.8$ & $19.2\pm1.7$\cr
    Var-CNN$^{*}$ &$7.5\pm0.4$  & $9.8\pm0.6$ & $11.5\pm1.0$ & $15.1\pm1.3$\cr
    % Var-CNN \cite{Bhat2019} &$0.8\pm0.3$  & $19.4\pm14.6$ & $50.5\pm7.0$ & $73.6\pm1.7$\cr
    \hline
    % DF+{\bf HDA} & \boldmath{ $75.9\pm0.4$} & $84.7\pm0.5$ & $87.9\pm0.7$& $90.4\pm0.4$\cr
    % Var-CNN+{\bf HDA} & $70.8\pm3.4$ & \boldmath{$85.7\pm0.6$} & \boldmath{$88.9\pm0.7$} & \boldmath{$91.7\pm0.6$}\cr
    ResNet-18+{\bf HDA} & $12.9\pm1.2$ & $27.9\pm2.8$ & $35.2\pm2.9$& $40.7\pm3.4$\cr
    ResNet-34+{\bf HDA} & $12.3\pm1.9$ & $28.1\pm3.7$ & $38.2\pm6.2$& $47.7\pm5.1$\cr
    Var-CNN+{\bf HDA} &$25.3\pm2.2$  & $46.9\pm1.9$ & $48.7\pm1.4$ & $63.2\pm1.8$\cr
    Var-CNN$^{*}$+{\bf HDA} & \boldmath{$26.0\pm4.7$} & \boldmath{$48.5\pm2.6$} & \boldmath{$59.7\pm1.9$} & \boldmath{$65.4\pm0.7$}\cr
    % DF \cite{Sirinam2018} & $25.4\pm2.3$ & $37.8\pm1.3$ & $48.0\pm1.8$& $49.3\pm2.1$\cr
    
    % Var-CNN \cite{Bhat2019} &$6.6\pm0.5$  & $9.7\pm0.3$ & $13.0\pm0.8$ & $19.6\pm1.2$\cr
    % \hline
    % DF+{\bf HDA} & {$27.8\pm2.1$} & $42.7\pm2.0$ & $53.4\pm1.2$& $58.8\pm0.4$\cr
    % Var-CNN+{\bf HDA} & \boldmath{$28.5\pm2.8$} & \boldmath{$49.8\pm1.8$} & \boldmath{$59.9\pm1.1$} & \boldmath{$65.2\pm1.3$}\cr
    \bottomrule
    \end{tabular}
    }
    % \end{threeparttable}
    % \label{table I}
\end{table}

\keypoint{Results. }
We reported the results of closed-world WF attack under 
WTF-PAD based defense in Table \ref{tab:result_defense}.
We made the following observations.
(1) Some hand-crafted feature based methods (CUMUL, AWF)
are superior over recent deep learning methods (ResNet-18, ResNet-34)
at the few-shot learning scenarios. This is mainly because
the latter suffers from lacking enough training samples,
resulting in model overfitting.
(2) Using our HDA for training data augmentation,
we can directly solve the data scarcity problem
and significantly boost the performances of previous deep learning methods.
As a result, both Var-CNN+HDA and Var-CNN$^{*}$+HDA outperform the other competitors by a large margin,
e.g., 4.6\% and 2.4\% gap over the best competitor CUMUL. 
(3) ResNet-34 is surpassed by Var-CNN and its variant dramatically. 
By benefiting more from our data augmentation, Var-CNN finally achieves the best results across all different shot cases.
This implies that Var-CNN has higher desire for large training data
with higher performance potential, as compared to ResNet-34.

\subsection{Ablation Studies}

We carried out a set of component analysis experiments 
to examine the exact effect of different designs of our method (HDA).
We adopted the most common closed-world attack scenario
{\em without} defense on the CW$_\text{100}$ dataset,
following the same setting as Section \ref{subsec:closed_world_exp}. 
 It is noteworthy that this dataset CW$_\text{100}$ is different from the dataset in Section \ref{subsec:closed_world_exp} because they are different subsets.
In this section, we evaluated the 15-shot learning case in particular,
using Var-CNN \cite{Bhat2019} as the deep learning model backbone.

\renewcommand{\arraystretch}{1.5} %控制行高
\begin{table} [h] %[!pb]
  \centering
  \caption{Effect of individual augmentation operations.
  }
  \label{tab:individual_operation}
    \begin{tabular}{l | r}
    \toprule
    Augmentation Operation    & Accuracy\cr
    \midrule
    \em None & $75.0\pm3.0$\cr
    \hline
    Random Rotation & {$92.4\pm0.3$}\cr
    Random Masking &  {$92.4\pm0.8$}\cr
    Random Mixing & $86.7\pm0.6$\cr
    Random Rotation + Masking & $92.7\pm0.4$ \cr
    Random Rotation + Mixing & $92.6\pm0.7$ \cr
    Random Masking + Mixing & $93.4\pm0.7$ \cr
    Random Rotation + Masking + Mixing (HDA) &
    \boldmath{$93.5\pm0.4$}\cr
    \bottomrule
    \end{tabular}
\end{table}

\keypoint{Individual augmentation operations. }
Recall that our data augmentation method (HDA) consists
of three different operations (random rotation, masking, and mixing).
We have demonstrated their performance advantages of them as a whole in 
varying test settings above.
For in-depth insights, examining their individual contributions
would be informative and necessary, as well as different combinations.
We conducted this experiments with an exhaustive set of operation 
combinations and reported the results in Table \ref{tab:individual_operation}.

It is observed that: 
(1)
Each of the three operations makes significant difference
in performance, with rotation and masking the best individual
operations that improve the classification accuracy by 17.4\%.
(2)
When jointly using any two augmentation operations,
the performance can be further increased.
The combination of masking and mixing gives the highest accuracy among them.
(3)
Combining all the three operations (HDA) achieves the best result with smaller deviation.
This suggests that all different operations are complementary and compatible to each other.

\keypoint{Augmentation optimization. }
For optimal data augmentation,
we propose a sequential optimization strategy (see Alg. \ref{alg}) for capturing the inter-dependence between
different augmentation operations applied.
To evaluate its effect, we compared with a baseline algorithm
that {\em independently} optimizes each augmentation parameter.

\renewcommand{\arraystretch}{1.5} %控制行高
\begin{table} [h] %[!tp]
  \centering
  \caption{
  Effect of augmentation optimization.
  }
  \label{tab:aug_opt}
    \begin{tabular}{c|c}
    \toprule
    Augmentation Optimization & Accuracy \cr
    \midrule
    % None & $75.0\pm3.0$\cr
    % RR, RE, Mixup&Direct Combination
    Independent & $92.1\pm0.5$\cr
    % RR, RE, Mixup&WF-Scalable&
    Sequential ({\bf Ours}) & 
    \boldmath{$93.5\pm0.4$} \cr
    \bottomrule
    \end{tabular}
\end{table}

As shown in Table \ref{tab:aug_opt}, 
the proposed optimization algorithm (see Alg. \ref{alg}) is clearly superior,
validating our consideration that there exist inter-dependence
between different augmentation operations when applied jointly on the same samples.
Note that we obtained this performance gain 
at the same cost as the baseline counterpart.
Besides, it is worth noting that even with the simpler optimization, our data augmentation method (HDA) can still greatly improve the previous deep learning model Var-CNN,
and achieve new state-of-the-art results (Table \ref{tab:aug_opt} vs. Table \ref{tab:cls_world_CW}).
This further validates that the proposed augmentation operations
are highly compatible with one another and can be applied together well.

\section{Conclusion}

We presented a  model-agnostic, simple yet surprisingly effective data augmentation method, called HDA, for few-shot website fingerprinting attack. This is an under-studied and realistically critical problem, as in practice only a handful of training samples per website can be feasibly collected due to the inherent high dynamics of internet networks and expensive label collection cost.
Importantly, we focus on deep learning based methods,
a line of new research efforts with vast potentials
for future investigations.
In particular, our HDA method offers three different
data augmentation operations,
including random rotation, masking,
and mixing in intra-sample and 
inter-sample fashion.
They can be applied to the same training samples
harmoniously with high complement
and compatibility.
Moreover, we introduce a sequential augmentation parameter
optimization method that captures the inter-dependence nature
between different operations when applied jointly.
With recent state-of-the-art deep learning WF attack models,
we conducted extensive experiments on four benchmark datasets
to validate the efficacy of our HDA method in both closed-world and 
open-world scenarios, with and without defense.
The results show that the proposed data augmentation method makes dramatic differences in performance and enables previous deep learning methods to outperform hand-crafted feature based counterparts in the few-shot learning setting
for the first time, often by a large margin.
This is achieved without making any artificial assumptions
of relevant, large auxiliary training data for model pre-training.
With our HDA method, collecting large training data frequently is eliminated, whilst still achieving stronger and more robust WF attack.
Finally, we performed detailed component analysis to 
diagnose the effect of individual model 
components.

% \section{Conclusion}
% The conclusion goes here.

% % if have a single appendix:
% %\appendix[Proof of the Zonklar Equations]
% % or
% %\appendix  % for no appendix heading
% % do not use \section anymore after \appendix, only \section*
% % is possibly needed

% % use appendices with more than one appendix
% % then use \section to start each appendix
% % you must declare a \section before using any
% % \subsection or using \label (\appendices by itself
% % starts a section numbered zero.)
% %

% \appendices

% % you can choose not to have a title for an appendix
% % if you want by leaving the argument blank
% \section{}

% \section*{Acknowledgment}


\ifCLASSOPTIONcaptionsoff
  \newpage
\fi



%\bibliographystyle{IEEEtran}


% \end{thebibliography}
\bibliographystyle{IEEEtranS.bst}
\bibliography{IEEEabrv.bib}

% \end{IEEEbiography}


% % insert where needed to balance the two columns on the last page with
% %\newpage



%\vfill


\end{document}